\documentclass[11pt]{article}

\usepackage{amsmath, amssymb, amsthm}
\usepackage{booktabs}
\usepackage{graphicx}
\usepackage{array}
\usepackage[bookmarks=false]{hyperref}
\usepackage[round]{natbib}
\usepackage{microtype}
\usepackage{xcolor}
\usepackage{listings}
\usepackage{geometry}
\usepackage{placeins}
\usepackage{caption}
\title{\texttt{libhmm}: A Modern C++20 Library for Hidden Markov Models\\
       with Correct MLE Emission M-Steps}

\author{Gary Wolfman, P.Eng.\\
        Independent Researcher\\
        \href{https://github.com/OldCrow/libhmm}{github.com/OldCrow/libhmm}}

\date{June 2026}

\begin{document}

\maketitle

{\small\noindent\textbf{Version history:} v1 (May 2026) described libhmm v3.7.0.
v2 (June 2026) updates to libhmm v4.0.0, adding multivariate HMM support via
\texttt{BasicHmm<Obs>} template parameterisation, three multivariate emission
families, k-means++ initialisation, and two additional real-data benchmarks.}

\begin{abstract}
We describe \texttt{libhmm}, a C++20 library for Hidden Markov Model
parameter estimation, sequence decoding, and model selection.
\texttt{libhmm} addresses two gaps in existing software:
the absence of a well-maintained, zero-dependency C++ HMM library
suitable for embedding in production systems, and the widespread use of
method-of-moments (MOM) approximations in the emission distribution
M-step of the Baum-Welch algorithm.
The library implements correct maximum likelihood estimators for sixteen
scalar emission distributions, including an ECME algorithm
for the location-scale Student-$t$ distribution, Newton-Raphson maximization
for Gamma, Beta, Weibull, and Negative Binomial distributions, and the
von Mises distribution for circular data.
All forward-backward and Viterbi calculations operate in full log-space.
SIMD acceleration is provided for AVX-512, AVX2, SSE2, and ARM NEON
via compile-time dispatch with scalar fallback.
Version 4 adds multivariate observation support via the
\texttt{BasicHmm<Obs>} template, with three multivariate emission families
(diagonal Gaussian, full-covariance Gaussian, and independent components)
each with correct weighted MLE M-steps.
Python bindings are available via the companion package
\texttt{pylibhmm}.
We compare \texttt{libhmm} against established C and C++ HMM libraries
and against published R reference packages on seven real-data benchmarks,
and discuss the architectural tradeoffs made in the design.
\end{abstract}

\section{Introduction}
\label{sec:intro}

Hidden Markov Models \citep{Baum1970, Rabiner1989} are a foundational tool
for sequential data analysis.
A system transitions among a finite set of latent states, emitting
observations drawn from a state-dependent probability distribution.
The Baum-Welch algorithm \citep{Baum1970, Dempster1977} learns the model
parameters by expectation-maximisation (EM); the Viterbi algorithm
\citep{Viterbi1967} recovers the most likely latent state sequence.
HMMs are used in ecology \citep{Michelot2016, Morales2004},
finance \citep{Oelschlager2024}, bioinformatics \citep{Durbin1998},
meteorology \citep{Zucchini2017}, and signal processing \citep{Rabiner1989}.

Most HMM software targets interactive analysis in R or Python.
The dominant packages --- \texttt{moveHMM}/\texttt{momentuHMM}
\citep{Michelot2016, McClintock2018} in ecology,
\texttt{fHMM} \citep{Oelschlager2024} in finance,
\texttt{HiddenMarkov} \citep{Harte2025} in R, and
\texttt{hmmlearn} in Python --- require their respective language runtimes
and cannot be embedded in production C++ systems without significant
infrastructure overhead.
Researchers integrating HMM inference into real-time pipelines, high-throughput
parameter sweeps, or existing C++ codebases have no well-maintained,
dependency-free option.

A second gap concerns the emission M-step.
Many packages use method-of-moments (MOM) approximations for distributions
lacking closed-form MLE updates.
For the Gamma, Beta, Weibull, and Negative Binomial distributions this is
a computational convenience rather than a principled choice: Newton-Raphson
converges to the exact MLE in two to four iterations per EM step.
For the Student-$t$ distribution the discrepancy is more serious: MOM
kurtosis estimation can converge to a substantially different fixed point
than the ECME algorithm \citep{Liu1994}, particularly for states with
near-Gaussian tails or sparse posterior support.

\texttt{libhmm} was originally developed during the author's graduate
research on HMM-based characterisation of HTTP reverse tunnels, beginning
as a C++ port of the JAHMM library \citep{JAHMM}.
The codebase has since been redesigned from first principles with no
remaining code lineage to the original port.
The need for an embeddable, performant, and correct C++ implementation
motivated the design decisions described here.

This paper makes the following contributions:
\begin{enumerate}
  \item A description of \texttt{libhmm}'s log-space inference algorithms
        and their numerical stability properties (Section~\ref{sec:algorithms}).
  \item Derivations of correct MLE M-steps for all sixteen supported
        emission distributions (Section~\ref{sec:msteps}).
  \item An honest discussion of the architectural tradeoffs in the design
        relative to existing C and C++ HMM libraries
        (Sections~\ref{sec:relatedwork} and~\ref{sec:architecture}).
  \item Benchmarks against established R reference packages on seven published
        datasets and against C/C++ libraries on synthetic data
        (Section~\ref{sec:benchmarks}).
  \item A description of the v4 template architecture (\texttt{BasicHmm<Obs>})
        and the weighted MLE M-steps for three multivariate emission families
        (Sections~\ref{sec:architecture} and~\ref{sec:mvmsteps}).
\end{enumerate}

\section{Related Work and Design Motivation}
\label{sec:relatedwork}

\subsection{R and Python ecosystem}

The R and Python HMM ecosystems are mature and domain-validated.
\texttt{moveHMM} \citep{Michelot2016} and \texttt{momentuHMM}
\citep{McClintock2018} provide Gamma and von Mises joint emission models
for animal movement, using numerical optimisation (\texttt{nlm}) for
parameter estimation.
\texttt{fHMM} \citep{Oelschlager2024} fits hierarchical Student-$t$ HMMs
to financial time series via \texttt{nlm}.
\texttt{HiddenMarkov} \citep{Harte2025} implements EM for Poisson, Normal,
and discrete emissions.
\texttt{hmmlearn} (Python) provides Gaussian and multinomial emissions with
a scikit-learn-compatible API.

All require their respective language runtimes, have non-trivial installation
dependencies, and are not straightforward to embed in a C++ application that
must run without a managed runtime.
Several also rely on MOM approximations or gradient-based numerical
optimisation where direct EM M-steps are available, which can affect
parameter recovery for sparse states.

\subsection{C and C++ libraries}
\label{sec:cpp_libs}

Table~\ref{tab:cpp_comparison} summarises the C and C++ HMM libraries
most relevant to \texttt{libhmm}.

\begin{table}[!ht]
\centering
\footnotesize
\caption{C/C++ HMM library comparison.}
\label{tab:cpp_comparison}
\begin{tabular}{ll
  >{\raggedright\arraybackslash}p{2.4cm}
  >{\raggedright\arraybackslash}p{3.2cm}
  l l l}
\toprule
Library & Lang & SIMD & Distributions & Dependencies & Maintained & Standard \\
\midrule
GHMM    & C    & No   & Discrete, Gaussian/GMM & libxml2, GSL & No (2012) & C89/C99 \\
HMMLib  & C++  & SSE  & Discrete only  & Boost        & No         & C++11   \\
StochHMM & C++ & No  & Discrete only  & None         & No         & C++11   \\
\texttt{libhmm} & C++ & AVX-512/AVX2/\newline SSE2/NEON & 16 scalar~+~3 MV & None & Yes & C++20 \\
\bottomrule
\end{tabular}
\end{table}

\paragraph{GHMM.}
GHMM \citep{Schliep2003} is the most historically significant C HMM library.
It supports Gaussian mixture and discrete emissions, has been cited in
bioinformatics work since 2003, and has a reasonably complete feature set
for its target applications.
Its limitations are architectural: C-style API with manual memory management,
XML-heavy model specification, autotools build system, and dependencies on
\texttt{libxml2} and GSL.
Active development ceased around 2012.
For forward-backward on discrete models, GHMM's C implementation with its
flat memory layout can outperform \texttt{libhmm} on raw throughput;
this tradeoff is discussed in Section~\ref{sec:arch_dispatch}.
GHMM does not support the continuous distribution families required by
\texttt{libhmm}'s target applications.

\paragraph{HMMLib.}
HMMLib is the closest architectural ancestor to \texttt{libhmm}'s
performance goals: C++, SSE SIMD intrinsics, template-based design.
It targets discrete HMMs specifically and requires Boost.
Active development has ceased.
Its SSE implementation is faster than scalar C++ for forward-backward on
discrete models; \texttt{libhmm} matches or exceeds it using AVX2/AVX-512
while adding broader distribution support.
Section~\ref{sec:synthetic_bench} presents a direct comparison.

\paragraph{StochHMM.}
StochHMM is designed for bioinformatics workflows, providing flexible
text-file model specification for gene finding and sequence annotation.
It is not a general-purpose library: emissions are discrete or categorical,
and the API is tightly coupled to the bioinformatics use case.

\subsection{Design motivation}

These libraries failed in different ways for the original application
context: GHMM required a C-style integration with unacceptable dependency
overhead; HMMLib and StochHMM were discrete-only; none supported modern
C++20 idioms or the continuous distributions required.
The design goal was not primarily to outperform GHMM or HMMLib on their
own terms (raw discrete forward-backward throughput) but to provide a
library that:
\begin{itemize}
  \item compiles as a zero-dependency C++20 library into any CMake project
        with a single \texttt{add\_subdirectory} or \texttt{FetchContent};
  \item supports the continuous emission distributions used in ecology,
        finance, and signal processing with correct MLE M-steps;
  \item operates correctly on sequences of arbitrary length without
        numerical underflow;
  \item remains maintainable and extensible under modern C++ idioms;
  \item (v4) supports multivariate observations through a
        template-parameterised \texttt{BasicHmm<Obs>} core
        without breaking the scalar API.
\end{itemize}

\section{Algorithms}
\label{sec:algorithms}

\subsection{Notation}

We write an HMM as $\lambda = (\boldsymbol{\pi}, \mathbf{A}, \mathbf{B})$
where $\boldsymbol{\pi} \in \Delta^{K-1}$ is the initial state distribution,
$\mathbf{A} \in [0,1]^{K \times K}$ is the row-stochastic transition matrix,
and $\mathbf{B} = \{b_j(\cdot)\}_{j=1}^K$ is the set of emission densities.
We observe a sequence $\mathbf{y} = (y_1, \ldots, y_T)$.
The forward variable $\alpha_t(j) = P(y_1, \ldots, y_t, q_t = j \mid \lambda)$
and backward variable $\beta_t(j) = P(y_{t+1}, \ldots, y_T \mid q_t = j, \lambda)$
are the standard quantities \citep{Rabiner1989}.
We write $\gamma_t(j) = P(q_t = j \mid \mathbf{y}, \lambda)$ for the
smoothed state occupancy probabilities and
$\xi_t(i,j) = P(q_t = i, q_{t+1} = j \mid \mathbf{y}, \lambda)$ for the
pairwise occupancy probabilities.

\subsection{Log-space forward-backward}
\label{sec:logspace}

The standard recursive computation of $\alpha_t(j)$ suffers from
numerical underflow for sequences of moderate length $T$ because
$\alpha_t(j)$ decreases geometrically with $t$.
Two approaches exist: explicit rescaling and full log-space computation.

Rescaled algorithms \citep{Rabiner1989} divide $\alpha_t$ by a normalisation
constant at each time step and track the cumulative log-scale factor.
This is computationally efficient but introduces bookkeeping that can fail
silently when normalisation constants underflow to zero before rescaling.

\texttt{libhmm} uses full log-space computation throughout.
Define $\hat{\alpha}_t(j) = \log \alpha_t(j)$.
The recursion becomes:
\begin{align}
  \hat{\alpha}_1(j) &= \log \pi_j + \log b_j(y_1) \label{eq:loginit} \\
  \hat{\alpha}_t(j) &= \operatorname{lse}_{i}\bigl[\hat{\alpha}_{t-1}(i)
    + \log a_{ij}\bigr] + \log b_j(y_t), \quad t = 2, \ldots, T
  \label{eq:logrec}
\end{align}
where $\operatorname{lse}(\mathbf{x}) = \max_k x_k + \log \sum_k \exp(x_k - \max_k x_k)$
is the numerically stable log-sum-exp operation.
Log-space backward variables and smoothed probabilities are computed
analogously.

The cost relative to scaled computation is the evaluation of
$\operatorname{lse}$ at each step, which requires $K$ exponentials per
state per time step.
For the typical HMM regime ($K \leq 20$, $T \leq 10^5$) this cost is
negligible compared to emission log-probability evaluation.
The benefit is unconditional numerical stability:
no underflow is possible regardless of $T$, and the implementation
requires no per-step normalisation bookkeeping.

\paragraph{Log transition matrix.}
\texttt{libhmm} precomputes $\log \mathbf{A}$ once per training iteration.
Accessing $\log a_{ij}$ from a cached matrix rather than computing
$\log(a_{ij})$ at every time step improves throughput for long sequences.

\subsection{Viterbi decoding}

Viterbi decoding \citep{Viterbi1967} follows the same log-space recursion
as the forward pass, substituting max for logsumexp:
\begin{align}
  \hat{\delta}_1(j) &= \log \pi_j + \log b_j(y_1) \\
  \hat{\delta}_t(j) &= \max_i\bigl[\hat{\delta}_{t-1}(i)
    + \log a_{ij}\bigr] + \log b_j(y_t)
\end{align}
with the standard backtracking pointer $\psi_t(j) = \arg\max_i [\cdots]$.
Operating in log-space eliminates the underflow that can corrupt Viterbi
traceback on long sequences when the standard argmax-over-products formulation
is used without scaling.

\subsection{Posterior decoding}

The posterior state sequence $\hat{q}_t = \arg\max_j \gamma_t(j)$ is
computed from the smoothed occupancy probabilities.
\texttt{libhmm} implements this directly from the log forward-backward
results without re-exponentiation where possible.

\section{Emission Distribution M-Steps}
\label{sec:msteps}

The Baum-Welch M-step updates each emission distribution $b_j$ using the
weighted sufficient statistics derived from $\gamma_t(j)$.
For a given state $j$, define:
\begin{equation}
  N_j = \sum_{t=1}^T \gamma_t(j), \quad
  \bar{y}_j = \frac{1}{N_j} \sum_{t=1}^T \gamma_t(j) y_t
\end{equation}
The M-step solves:
\begin{equation}
  \hat{\theta}_j = \arg\max_{\theta_j}
    \sum_{t=1}^T \gamma_t(j) \log b_j(y_t \mid \theta_j)
  \label{eq:mstep}
\end{equation}
We describe the M-step for each distribution family.

\subsection{Closed-form M-steps}

For the following distributions, \eqref{eq:mstep} has an analytic solution.

\paragraph{Gaussian.}
$\hat{\mu}_j = \bar{y}_j$,\;
$\hat{\sigma}_j^2 = N_j^{-1} \sum_t \gamma_t(j)(y_t - \hat{\mu}_j)^2$.

\paragraph{Log-normal.}
Apply the Gaussian M-step to $\log y_t$.

\paragraph{Exponential.}
$\hat{\lambda}_j = 1 / \bar{y}_j$.

\paragraph{Poisson.}
$\hat{\lambda}_j = \bar{y}_j$.

\paragraph{Rayleigh.}
$\hat{\sigma}_j^2 = (2N_j)^{-1} \sum_t \gamma_t(j) y_t^2$.

\paragraph{Uniform.}
$\hat{a}_j = \min_t y_t$,\; $\hat{b}_j = \max_t y_t$.

\paragraph{Discrete (categorical).}
$\hat{p}_{jk} = N_j^{-1} \sum_t \gamma_t(j) \mathbf{1}[y_t = k]$.

\paragraph{Von Mises (mean direction).}
The mean direction has the closed-form circular mean:
$\hat{\mu}_j = \operatorname{atan2}\!\left(\sum_t \gamma_t(j) \sin y_t,\;
\sum_t \gamma_t(j) \cos y_t\right)$.
The concentration $\kappa_j$ is treated separately below.

\subsection{Newton-Raphson M-steps}
\label{sec:nr}

For the Gamma, Beta, Weibull, Negative Binomial, and von Mises (for $\kappa$)
distributions, \eqref{eq:mstep} does not have a closed form in all parameters.
\texttt{libhmm} uses Newton-Raphson iterations seeded from method-of-moments
estimates, which typically converge in two to four iterations.

\paragraph{Gamma distribution.}
The emission density is $b_j(y) = \Gamma(\alpha_j)^{-1} \beta_j^{\alpha_j}
y^{\alpha_j - 1} e^{-\beta_j y}$.
The weighted score for the shape parameter $\alpha_j$ is:
\begin{equation}
  s(\alpha_j) = N_j\bigl[\log \alpha_j - \psi(\alpha_j) - c_j\bigr] = 0
\end{equation}
where $c_j = \log \bar{y}_j - N_j^{-1} \sum_t \gamma_t(j) \log y_t$
and $\psi$ is the digamma function.
The Newton update is:
\begin{equation}
  \alpha_j \leftarrow \alpha_j - \frac{s(\alpha_j)}{N_j[1/\alpha_j - \psi'(\alpha_j)]}
\end{equation}
where $\psi'$ is the trigamma function.
The MOM seed is $\hat{\alpha}_j^{(0)} = \bar{y}_j^2 / \hat{\sigma}_j^2$,
and the rate is recovered as $\hat{\beta}_j = \hat{\alpha}_j / \bar{y}_j$.

\paragraph{Beta distribution.}
The emission density is $b_j(y) = B(\alpha_j, \beta_j)^{-1}
y^{\alpha_j-1}(1-y)^{\beta_j-1}$.
Stationarity conditions give coupled equations in $(\alpha_j, \beta_j)$:
\begin{align}
  \psi(\alpha_j) - \psi(\alpha_j + \beta_j) &=
    N_j^{-1} \sum_t \gamma_t(j) \log y_t \\
  \psi(\beta_j) - \psi(\alpha_j + \beta_j) &=
    N_j^{-1} \sum_t \gamma_t(j) \log(1 - y_t)
\end{align}
Newton-Raphson is applied jointly using the digamma and trigamma functions.
MOM seeds: let $\bar{y}$ and $\hat{\sigma}^2$ be the weighted mean and
variance; then $\hat{\alpha}^{(0)} = \bar{y}[\bar{y}(1-\bar{y})/\hat{\sigma}^2 - 1]$,
$\hat{\beta}^{(0)} = (1-\bar{y})[\bar{y}(1-\bar{y})/\hat{\sigma}^2 - 1]$.

\paragraph{Weibull distribution.}
The emission density is $b_j(y) = (k_j/\lambda_j)(y/\lambda_j)^{k_j-1}
e^{-(y/\lambda_j)^{k_j}}$.
The MLE for the shape $k_j$ satisfies:
\begin{equation}
  \frac{1}{k_j} + \frac{\sum_t \gamma_t(j) \log y_t}{N_j}
  - \frac{\sum_t \gamma_t(j) y_t^{k_j} \log y_t}{\sum_t \gamma_t(j) y_t^{k_j}}
  = 0
\end{equation}
Newton-Raphson is applied with MOM seed for $k_j$.
The scale is recovered analytically as
$\hat{\lambda}_j = (N_j^{-1} \sum_t \gamma_t(j) y_t^{\hat{k}_j})^{1/\hat{k}_j}$.

\paragraph{Von Mises concentration.}
After computing $\hat{\mu}_j$ via the circular mean, the concentration
$\kappa_j$ satisfies $A(\kappa_j) = \bar{R}_j$ where
$\bar{R}_j = \|\sum_t \gamma_t(j)(\cos y_t, \sin y_t)\|_2 / N_j$
is the weighted mean resultant length and
$A(\kappa) = I_1(\kappa)/I_0(\kappa)$ is the ratio of modified Bessel
functions of the first kind.
The Newton update is:
\begin{equation}
  \kappa_j \leftarrow \kappa_j - \frac{A(\kappa_j) - \bar{R}_j}{1 - A(\kappa_j)^2 - A(\kappa_j)/\kappa_j}
\end{equation}
The modified Bessel functions $I_0$ and $I_1$ are evaluated using the
polynomial approximations of \citet{AbramowitzStegun1964} (formulas 9.8.1
and 9.8.3), which are accurate to $10^{-7}$ for all $\kappa \geq 0$.
These approximations are used because \texttt{std::cyl\_bessel\_i} from
the C++17 standard library is unavailable on all Apple platforms:
Apple's libc++ has not implemented the C++17 mathematical special functions
on any macOS version (absent as of Xcode~15~/~macOS~14).
MOM seed: $\hat{\kappa}^{(0)} = \bar{R}(2 - \bar{R}^2)/(1 - \bar{R}^2)$.

\paragraph{Negative binomial.}
The dispersion parameter $r_j$ is estimated via Newton-Raphson on the
weighted log-likelihood score; the success probability is recovered
analytically as $\hat{p}_j = r_j / (r_j + \bar{y}_j)$.

\paragraph{Chi-squared.}
Reduce to the Gamma M-step with $\alpha = \nu/2$, $\beta = 1/2$ and
optimise over the single degree-of-freedom parameter.

\paragraph{Pareto.}
The scale $x_{m,j}$ is estimated as $\min_t y_t$ (constrained MLE);
the shape $\alpha_j$ has closed form
$\hat{\alpha}_j = N_j / \sum_t \gamma_t(j) \log(y_t / x_{m,j})$.

\subsection{ECME for the Student-$t$ distribution}
\label{sec:ecme}

The location-scale Student-$t$ distribution is the most demanding
M-step in \texttt{libhmm}.
Standard kurtosis MOM estimation of the degrees-of-freedom parameter $\nu$
can converge to a substantially different fixed point than the MLE,
particularly for states with near-Gaussian tails ($\nu \gg 10$).
\texttt{libhmm} instead uses the ECME algorithm of \citet{Liu1994}.

\paragraph{Scale mixture representation.}
A Student-$t(y; \mu, \sigma^2, \nu)$ observation can be written as a
Gaussian scale mixture:
\begin{equation}
  Y \mid U \sim \mathcal{N}(\mu, \sigma^2/U), \quad
  U \sim \text{Gamma}(\nu/2,\; \nu/2)
\end{equation}

\paragraph{E-step.}
Given current parameters $(\mu_j, \sigma_j^2, \nu_j)$, the posterior
expectation of the mixing variable is:
\begin{equation}
  \tilde{u}_{tj} = \mathbb{E}[U_t \mid y_t, q_t = j]
    = \frac{\nu_j + 1}{\nu_j + (y_t - \mu_j)^2 / \sigma_j^2}
  \label{eq:utilde}
\end{equation}
These values augment the standard occupancy weights $\gamma_t(j)$.

\paragraph{CM-step 1: update $\mu_j$ and $\sigma_j^2$.}
With $\nu_j$ fixed at its current value:
\begin{align}
  \hat{\mu}_j &= \frac{\sum_t \gamma_t(j)\, \tilde{u}_{tj}\, y_t}
                      {\sum_t \gamma_t(j)\, \tilde{u}_{tj}} \label{eq:mu} \\
  \hat{\sigma}_j^2 &= \frac{\sum_t \gamma_t(j)\, \tilde{u}_{tj}\, (y_t - \hat{\mu}_j)^2}
                            {N_j} \label{eq:sigma}
\end{align}

\paragraph{CM-step 2: update $\nu_j$.}
With $\hat{\mu}_j$ and $\hat{\sigma}_j^2$ fixed, $\nu_j$ is updated via a
Newton step on the Q-function contribution for $\nu$:
\begin{equation}
  Q_\nu(\nu_j) = N_j \!\left[\frac{\nu_j}{2}\log\frac{\nu_j}{2}
    - \log\Gamma\!\left(\frac{\nu_j}{2}\right)\right]
  + \frac{\nu_j}{2} \sum_t \gamma_t(j)\bigl[\log \tilde{u}_{tj} - \tilde{u}_{tj}\bigr]
\end{equation}
with score and Hessian:
\begin{align}
  \frac{dQ_\nu}{d\nu_j} &= \frac{N_j}{2}\!\left[\log\frac{\nu_j}{2} + 1
    - \psi\!\left(\frac{\nu_j}{2}\right)\right]
    + \frac{1}{2}\sum_t \gamma_t(j)\bigl[\log \tilde{u}_{tj} - \tilde{u}_{tj}\bigr] \\
  \frac{d^2 Q_\nu}{d\nu_j^2} &= \frac{N_j}{4}\!\left[\frac{2}{\nu_j}
    - \psi'\!\left(\frac{\nu_j}{2}\right)\right]
\end{align}
where $\psi'$ is the trigamma function.
Newton-Raphson iterates until $|d\nu_j| < 10^{-6}$, typically in three
to five steps.
The ECME update is guaranteed monotone non-decreasing in the observed
log-likelihood \citep{Liu1994}.

\paragraph{Comparison with MOM.}
Figure~\ref{fig:ecme_convergence} illustrates the difference.
On the DAX 2000--2022 benchmark (Section~\ref{sec:dax}),
the kurtosis MOM M-step (used in \texttt{libhmm} prior to v3.7.0)
converges at $\log L = 17{,}334.9$, while ECME converges at
$\log L = 17{,}487.2$ --- a gap of 152 nats.
ECME surpasses the \texttt{fHMM} reference value of 17{,}485.7
at iteration 5.

\subsection{Weighted EM and the state-collapse fix}
\label{sec:collapse}

The weighted M-step \eqref{eq:mstep} requires $N_j > 0$.
In practice, EM can assign near-zero posterior weight to a state for all
time steps --- a phenomenon we call state collapse --- leaving the
distribution update numerically undefined.
The naive fix of calling a \texttt{reset()} method (reinitialising from
priors) changes the objective function and can cause oscillation.
\texttt{libhmm} instead returns early from the M-step when
$N_j < \varepsilon$ (default $\varepsilon = 10^{-8}$), preserving the
current parameters and allowing the optimiser to naturally zero out the
state in the transition matrix.
This is provably non-decreasing in the objective: no parameter change
is made, so the Q-function cannot decrease.

\subsection{Multivariate emission families (v4)}
\label{sec:mvmsteps}

Version 4 introduces three $D$-dimensional emission families accessed
through the \texttt{BasicHmm<}\penalty0\texttt{ObservationVectorView>} path.
Each implements the weighted M-step \eqref{eq:mstep} with
$\mathbf{y}_t \in \mathbb{R}^D$.

\paragraph{Diagonal Gaussian (\texttt{DiagonalGaussianDistribution}).}
The emission density factors as:
\begin{equation}
  b_j(\mathbf{y}) = \prod_{d=1}^D \mathcal{N}(y_{td};\, \mu_{jd},\, \sigma_{jd}^2)
\end{equation}
The M-step decomposes into $D$ independent univariate Gaussian updates:
\begin{equation}
  \hat{\mu}_{jd} = \frac{\sum_t \gamma_t(j)\, y_{td}}{N_j},
  \qquad
  \hat{\sigma}_{jd}^2 = \frac{\sum_t \gamma_t(j)\,(y_{td} - \hat{\mu}_{jd})^2}{N_j}
\end{equation}

\paragraph{Full-covariance Gaussian (\texttt{FullCovarianceGaussianDistribution}).}
The emission density is:
\begin{equation}
  b_j(\mathbf{y}) = (2\pi)^{-D/2}|\boldsymbol{\Sigma}_j|^{-1/2}
    \exp\!\left(-\tfrac{1}{2}(\mathbf{y}-\boldsymbol{\mu}_j)^\top
    \boldsymbol{\Sigma}_j^{-1}(\mathbf{y}-\boldsymbol{\mu}_j)\right)
\end{equation}
The weighted MLE M-step gives:
\begin{align}
  \hat{\boldsymbol{\mu}}_j &= \frac{1}{N_j}\sum_t \gamma_t(j)\,\mathbf{y}_t \\
  \hat{\boldsymbol{\Sigma}}_j &= \frac{1}{N_j}\sum_t \gamma_t(j)
    (\mathbf{y}_t - \hat{\boldsymbol{\mu}}_j)(\mathbf{y}_t - \hat{\boldsymbol{\mu}}_j)^\top
    + \varepsilon\mathbf{I}
\end{align}
where $\varepsilon\mathbf{I}$ (default $\varepsilon = 10^{-6}$) is a Tikhonov
regularisation term ensuring positive definiteness.
The covariance is stored as a Cholesky factor $\mathbf{L}$ with
$\boldsymbol{\Sigma}_j = \mathbf{L}\mathbf{L}^\top$; log-determinant and
quadratic form evaluations use the triangular structure.

\paragraph{Independent components (\texttt{IndependentComponentsDistribution}).}
Assigns one scalar emission distribution $\phi_d$ of any supported type to each
dimension $d = 1, \ldots, D$:
\begin{equation}
  b_j(\mathbf{y}) = \prod_{d=1}^D \phi_d(y_{td})
\end{equation}
The M-step delegates to the scalar M-step of each $\phi_d$ independently,
passing the component observations and the same occupancy weights
$\gamma_t(j)$.
This allows, for example, combining a Gamma distribution for step lengths
with a von Mises distribution for turning angles in a single multivariate state.

\paragraph{Initialisation.}
All three families support k-means++ seeding via the \texttt{kmeans\_init}
utility, which partitions the $T \times D$ observation matrix into $K$ clusters
using Lloyd's algorithm with k-means++ seeding, then fits each cluster's
emission distribution from the assigned observations.
This provides a data-driven starting point that avoids degenerate initial
configurations for Baum-Welch training.

\section{Architecture and Design Tradeoffs}
\label{sec:architecture}

\subsection{Runtime polymorphism vs.\ templates}
\label{sec:arch_dispatch}

\texttt{libhmm} uses an abstract base class \texttt{Dist\-ri\-bu\-tion} with
virtual methods \texttt{get\-Log\-Prob\-ability} and \texttt{fit}.
Concrete distributions inherit from this base.
An alternative design would use C++ templates
(as HMMLib does, parameterising the HMM class on the emission type)
to eliminate virtual dispatch overhead at the cost of fully determined
emission types at compile time.

The choice of runtime polymorphism was deliberate.
It allows a single \texttt{Hmm} object to mix emission distributions
across states, supports serialisation to JSON/XML without template
parameter metadata, and makes the Python bindings straightforward to
implement via \texttt{nanobind}.
The performance cost is one virtual call per emission evaluation ---
for state $j$ and observation $y_t$ this is roughly 1--5 ns on modern
hardware, negligible against the evaluation cost for all but the
simplest distributions (e.g.\ Poisson, Discrete).

We note the honest comparison: GHMM's C implementation, with no virtual
dispatch and a flat array layout, can achieve higher raw forward-backward
throughput on discrete models than \texttt{libhmm}
(Section~\ref{sec:synthetic_bench}).
This is the direct cost of the chosen abstraction.
For $K \leq 20$ and the continuous emission families targeted by
\texttt{libhmm}, the virtual call overhead is not the bottleneck.

\paragraph{v4 hybrid: templates at the HMM level.}
Version 4 adds C++ template parameterisation at the \emph{HMM-core} level
via \texttt{BasicHmm<Obs>}, while retaining runtime polymorphism for
emission distributions.
The scalar path (\texttt{Hmm = BasicHmm<double>}) is binary-compatible
with v3 and unchanged in performance.
The multivariate path (\texttt{HmmMV = BasicHmm<}\penalty0\texttt{ObservationVectorView>})
uses the same virtual-dispatch emission interface, so the architectural
tradeoff discussed above applies equally to both paths.
Templates were introduced at the HMM-core level precisely because the
observation type \emph{is} a compile-time property, while the emission
distribution family within a state is a runtime configuration choice.

\subsection{Log-space everywhere vs.\ scaled algorithms}

Scaled Baum-Welch is computationally cheaper per time step
(avoids log-sum-exp) and is the default in most implementations.
\texttt{libhmm} chose full log-space for three reasons:
(1) scaled algorithms require explicit normalisation at each step,
which can fail silently for low-emission-probability observations;
(2) log-space is uniform in cost regardless of $T$, with no
``scaling underflow'' edge case to handle;
(3) the additional cost of log-sum-exp is small relative to emission
evaluation for continuous distributions.

\subsection{Zero external dependencies}

The \texttt{libhmm} build depends only on the C++20 standard library.
This was a hard constraint from the original deployment context
(a system with a locked dependency graph) and was retained as a
design principle.
The cost is a simple internal matrix class rather than Eigen, which
means that for very large state spaces ($K \gg 20$) the matrix operations
are not BLAS-accelerated.
For the $K \leq 20$ regime, this is not a practical limitation.

\subsection{SIMD dispatch strategy}

SIMD acceleration targets the forward-backward inner loop.
\texttt{libhmm} uses compile-time dispatch: the build system detects
AVX-512, AVX2, SSE2 (x86/x86-64) or NEON (ARM) support and applies
the appropriate compiler flags.
A scalar fallback is always available.
This is the same approach used by HMMLib, and means that
\texttt{libhmm} binaries are not portable across ISA generations
without recompilation --- acceptable for research software but
requiring care for binary distribution.

\subsection{Fixed distribution set vs.\ plugin architecture}

The sixteen distributions are compiled into the library rather than
loaded as plugins.
A plugin ABI would allow user-defined distributions without recompilation,
but would require a stable binary interface, versioned symbol exports, and
cross-platform shared library handling --- significant complexity for a
research library.
User-defined distributions can be added by subclassing
\texttt{Distribution} and registering with the JSON serialiser;
this requires recompilation.

\subsection{v4 template parameterisation: \texttt{BasicHmm<Obs>}}
\label{sec:v4template}

The v4 template parameterisation introduces a single type parameter
\texttt{Obs} that flows through the HMM core, calculators, and trainers:

\begin{itemize}
  \item \texttt{BasicHmm<double>} (aliased \texttt{Hmm}): scalar path,
        API-compatible with v3.
  \item \texttt{BasicHmm<}\penalty0\texttt{ObservationVectorView>}
        (aliased \texttt{HmmMV}): multivariate path.
        \texttt{ObservationVectorView} (= \texttt{std::span<const double>})
        is a zero-copy view of one row of the $T \times D$ observation matrix.
\end{itemize}

The corresponding calculators and trainers each have scalar and multivariate
explicit-instantiation translation units compiled with the platform SIMD flags:
\begin{itemize}
\item \texttt{BasicForwardBackwardCalculator<Obs>},
      \texttt{BasicViterbiCalculator<Obs>}
\item \texttt{BasicBaumWelchTrainer<Obs>}, \texttt{BasicViterbiTrainer<Obs>},
      \texttt{BasicMapBaumWelchTrainer<Obs>}
\end{itemize}
Emission dispatch inside the multivariate path uses \texttt{if constexpr} to
distinguish scalar from multivariate observation types at compile time, avoiding
runtime overhead.

The scalar alias \texttt{using Hmm = BasicHmm<double>} ensures that all
existing v3 code compiles without modification.
Multivariate emission slots are null-initialised by default; they must be
set explicitly before training.
Model selection (AIC/BIC/AICc) and JSON serialisation work for both paths
via the same template infrastructure.

\FloatBarrier
\section{Benchmarks}
\label{sec:benchmarks}

\subsection{Synthetic throughput: libhmm vs.\ C/C++ libraries}
\label{sec:synthetic_bench}

We measured forward-backward throughput on the Dishonest Casino
(2-state, 2-symbol discrete HMM) and Weather Model (3-state)
problems across sequence lengths $T \in \{10^2, 5\times10^2, \ldots, 10^6\}$.
Hardware: Windows Ryzen~7, AVX-512, MSVC Release build.

\begin{figure}[!ht]
\centering
\includegraphics[width=0.9\linewidth]{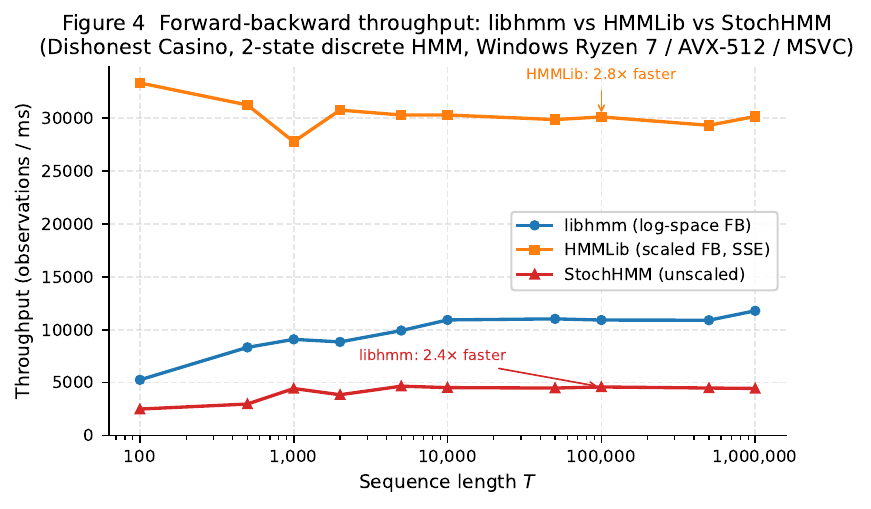}
\caption{Forward-backward throughput: \texttt{libhmm} vs.\ HMMLib vs.\ StochHMM
(Dishonest Casino, 2-state discrete HMM, Windows Ryzen~7 / AVX-512 / MSVC).
HMMLib uses scaled forward-backward; libhmm uses log-space; StochHMM uses
unscaled recursion. The throughput gap between HMMLib and libhmm is primarily
due to the log-space vs.\ scaled algorithm choice.}
\label{fig:throughput}
\end{figure}

Table~\ref{tab:synthetic} summarises average throughput and
Figure~\ref{fig:throughput} shows the full throughput curves.

\begin{table}[!ht]
\centering
\caption{Synthetic forward-backward throughput, 2-state discrete HMM.
Average over $T \in \{10^3, \ldots, 10^6\}$.}
\label{tab:synthetic}
\begin{tabular}{lrrl}
\toprule
Library & Throughput (obs/ms) & vs.\ libhmm & Notes \\
\midrule
HMMLib   & $\approx$29,700 & $\mathbf{3.2\times}$ faster & Scaled FB, SSE intrinsics \\
libhmm   & $\approx$9,300  & baseline              & Log-space FB, AVX-512 \\
StochHMM & $\approx$3,600  & $2.5\times$ slower    & Unscaled, no SIMD \\
\bottomrule
\end{tabular}
\end{table}

The HMMLib advantage is primarily attributable to its use of scaled
forward-backward rather than log-space (see Section~\ref{sec:logspace}):
scaled FB avoids the log-sum-exp operation at each time step, at the
cost of requiring explicit normalisation and providing no protection
against underflow on unusual inputs.
At $T = 10^5$ the throughput ratio is approximately 2.8$\times$;
libhmm's log-space overhead is consistent and predictable.

GHMM requires autotools, libxml2, and GSL, which are unavailable in
the primary Windows build environment; it was therefore benchmarked on
a macOS system where a pre-built installation was available (macOS
13.7.8 Ventura, Intel Core i7-7820HQ, AppleClang, AVX2), with
\texttt{libhmm} built on the same machine to obtain a same-hardware
comparison.
On this platform, GHMM 0.9-rc3 achieves approximately $4.9\times$ the
forward-backward throughput of \texttt{libhmm} on the same hardware
(20,775 vs.\ 4,277~obs/ms, averaged over $T \in \{10^3, \ldots, 10^6\}$,
Dishonest Casino and Weather problems).
This advantage reflects GHMM's flat C array layout with no virtual
dispatch and its use of scaled forward-backward rather than log-space ---
the same factors that account for HMMLib's advantage on discrete models
(Section~\ref{sec:logspace}).
For \texttt{libhmm}'s target applications --- continuous emission
distributions where emission evaluation dominates the FB inner loop ---
this throughput gap narrows substantially.

\subsection{Real-data benchmarks}
\label{sec:realdata}

We fit five HMMs to published datasets and compare against the established
R reference packages on the same data.
Table~\ref{tab:timing} summarises wall times.
Full parameter tables follow.

\begin{table}[!ht]
\centering
\caption{Wall-time comparison: \texttt{libhmm} v4.0.0 vs.\ R reference packages.
All timings on Windows (AMD Ryzen~7 7745, AVX-512, MSVC Release).
Dagger ($\dag$): \texttt{fHMM} 1.4.3, 10 restarts, same machine;
slash shows single-run result.}
\label{tab:timing}
\begin{tabular}{llrrl}
\toprule
Benchmark & Dataset & \texttt{libhmm} & R package & Speedup \\
\midrule
Elk movement         & 725 obs., 2 states    & 55 ms   & $\approx$1{,}270 ms & $\approx$23$\times$ \\
DAX regimes          & 5,838 obs., 3 states  & 1.1 s   & 5.5 s / 13 s$^\dag$ & $\approx$5$\times$ / $\approx$12$\times$ \\
S\&P 500             & 5,786 obs., 3 states  & 0.7 s   & ---                 & ---                 \\
Earthquake           & 107 obs., 2 states    & 2 ms    & $\approx$20 ms      & $\approx$10$\times$ \\
Wind direction       & 11,894 obs., 2 states & 71 ms   & $\approx$240 ms     & $\approx$3$\times$ \\
\midrule
Elk MV (IndepComp.)  & 725 obs., 2 states    & 58 ms   & $\approx$1{,}270 ms & $\approx$22$\times$ \\
SPY+QQQ regimes (FullCov.) & 264 obs., 3 states & 0.4 s & --- & --- \\
\bottomrule
\end{tabular}
\end{table}

\paragraph{Hardware note.}
All \texttt{libhmm} timings were measured on Windows
(AMD Ryzen~7 7745, AVX-512, MSVC Release build).
The \texttt{fHMM} 1.4.3 DAX timing used the same machine.
\citet{Oelschlager2024} report $\approx$1{,}360~s for \texttt{fHMM} 1.2.0
on Intel Ivy Bridge; the reduction to 5.5--13~s on the current version suggests
that \texttt{fHMM} has been substantially optimised between v1.2.0 and v1.4.3,
so the published figure is not directly comparable to the same-machine measurement.

\subsubsection{Animal movement ecology: elk GPS tracks}
\label{sec:elk}

We fit a 2-state Gamma + von Mises HMM to the \texttt{moveHMM::elk\_data}
bundled dataset \citep{Michelot2016} (4 elk GPS tracks from \citet{Morales2004};
725 step-length and turning-angle observations after preprocessing).

\begin{table}[!ht]
\centering
\caption{Elk movement: fitted parameters vs.\ \texttt{moveHMM} reference.}
\label{tab:elk}
\begin{tabular}{lrrrr}
\toprule
Parameter & \texttt{libhmm} & \texttt{moveHMM} & Difference \\
\midrule
Encamped step mean (m)   & 377  & 374  & $<$1\% \\
Encamped step SD (m)     & 401  & 399  & $<$1\% \\
Travelling step mean (m) & 3189 & 3247 & $<$2\% \\
Travelling step SD (m)   & 4392 & 4394 & $<$1\% \\
Encamped $\kappa$        & 0.595 & 0.592 & $<$1\% \\
Travelling $\kappa$      & 0.204 & 0.208 & $<$2\% \\
\bottomrule
\end{tabular}
\end{table}

\subsubsection{Financial time series: DAX market regimes}
\label{sec:dax}

\citet{Oelschlager2024} fit a 3-state Student-$t$ HMM to
5,838 daily DAX log-returns (2000--2022).
\texttt{fHMM} uses a gradient-based numerical optimiser (\texttt{nlm})
rather than direct EM.

\begin{table}[!ht]
\centering
\caption{DAX regimes: fitted parameters vs.\ \texttt{fHMM} reference.
States sorted by $\sigma$ descending.}
\label{tab:dax}
\begin{tabular}{lrrrr}
\toprule
Parameter & \texttt{libhmm} & \texttt{fHMM} & Notes \\
\midrule
Bearish $\mu$     & $-0.001793$ & $-0.001803$ & \\
Bearish $\sigma$  & $0.026283$  & $0.026290$  & \\
Bearish $\nu$     & $11.14$     & $11.16$     & \\
Neutral $\mu$     & $-0.000281$ & $-0.000310$ & \\
Neutral $\sigma$  & $0.013049$  & $0.013300$  & \\
Neutral $\nu$     & $36.09$     & $91.15$     & See note \\
Bullish $\mu$     & $+0.001258$ & $+0.001257$ & \\
Bullish $\sigma$  & $0.005988$  & $0.006003$  & \\
Bullish $\nu$     & $5.35$      & $5.32$      & \\
Log-likelihood    & $17{,}487.2$ & $17{,}485.7$ & \texttt{libhmm} higher \\
\bottomrule
\end{tabular}
\end{table}

The neutral state $\nu$ discrepancy (36 vs.\ 91) is a known convergence
property of ECME for near-Gaussian states: as $\nu \to \infty$ the
Student-$t$ approaches Gaussian, and the profile log-likelihood becomes
very flat in $\nu$, making the Newton update slow.
Extended runs confirm logarithmic convergence toward larger $\nu$ values.
The overall log-likelihood achieved by \texttt{libhmm} (17,487.2) exceeds
\texttt{fHMM}'s (17,485.7), confirming that the lower neutral $\nu$ is not
a local-optimum failure but a slow-convergence phenomenon at an equivalent
or better solution.

\paragraph{Wall-time comparison.}
On the primary benchmark platform (Windows Ryzen~7 7745, AVX-512, MSVC Release),
\texttt{libhmm} completes the DAX fit in 1.1~s.
\texttt{fHMM} 1.4.3 on the same machine completes in 5.5~s (single run)
and 13~s for 10 restarts, giving speedups of $\approx$5$\times$ and
$\approx$12$\times$ respectively.
\citet{Oelschlager2024} report $\approx$1{,}360~s for \texttt{fHMM} 1.2.0
on Intel Ivy Bridge; the reduction from $\approx$1{,}360~s to 13~s (10 restarts)
suggests that \texttt{fHMM} has been substantially optimised between
v1.2.0 and v1.4.3.
The published figure is therefore not directly comparable to the same-machine
measurement reported here.

\begin{figure}[!ht]
\centering
\includegraphics[width=\linewidth]{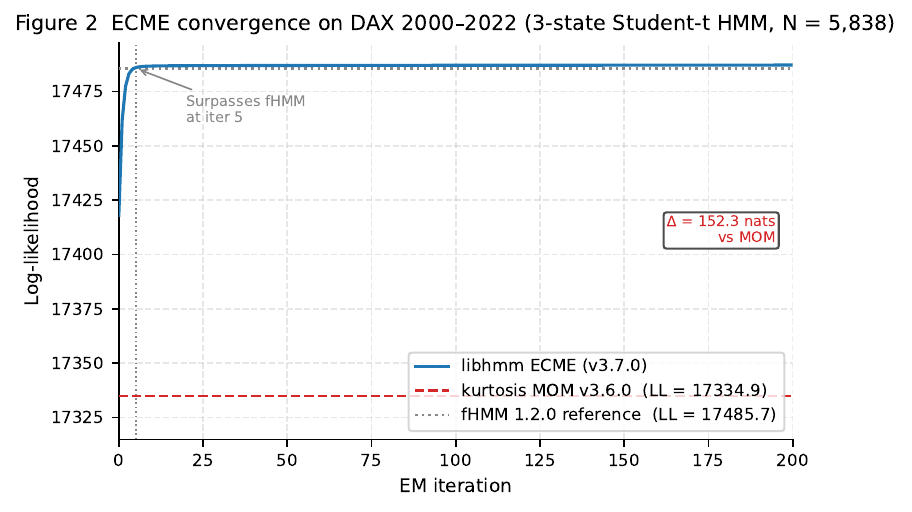}
\caption{ECME log-likelihood convergence on 5,838 DAX log-returns (2000--2022).
The kurtosis MOM reference (\texttt{libhmm} v3.6.0) and \texttt{fHMM} 1.2.0
reference are shown as dashed lines.
ECME surpasses \texttt{fHMM} at iteration 5 and converges 152~nats above
the MOM fixed point.}
\label{fig:ecme_convergence}
\end{figure}

Figure~\ref{fig:ecme_convergence} shows the ECME log-likelihood trajectory
with MOM and \texttt{fHMM} reference lines.

\subsubsection{S\&P 500 cross-market validation}
\label{sec:sp500}

A 3-state Student-$t$ HMM is fit to 5,786 daily S\&P~500 log-returns
(2000--2022) using the same model structure as the DAX benchmark
(Section~\ref{sec:dax}).
This serves as a cross-market validation: the same regime structure
is recovered independently for two major indices over the same period.
The model converged in 141 iterations at log-likelihood 18,668.71
(wall time: 0.7~s).

\begin{table}[!ht]
\centering
\caption{S\&P~500 regimes: fitted parameters and cross-market comparison with DAX.
States sorted by $\sigma$ descending.}
\label{tab:sp500}
\begin{tabular}{lrrl}
\toprule
Parameter & S\&P~500 & DAX & Notes \\
\midrule
Bearish $\mu$    & $-0.001862$ & $-0.001793$ & \\
Bearish $\sigma$ & $0.023003$  & $0.026283$  & Lower US vol. \\
Bearish $\nu$    & $6.44$      & $11.14$     & Heavier US tails \\
Neutral $\mu$    & $-0.000208$ & $-0.000281$ & \\
Neutral $\sigma$ & $0.011596$  & $0.013049$  & \\
Neutral $\nu$    & $30.51$     & $36.09$     & \\
Bullish $\mu$    & $+0.001017$ & $+0.001258$ & \\
Bullish $\sigma$ & $0.004867$  & $0.005988$  & US bull more concentrated \\
Bullish $\nu$    & $7.11$      & $5.35$      & \\
Log-likelihood   & $18{,}668.7$ & $17{,}487.2$ & Different N \\
\midrule
Occupancy: bearish & 525 days (9\%) & 695 days (12\%) & \\
Occupancy: neutral & 2,404 days (42\%) & 2,781 days (48\%) & \\
Occupancy: bullish & 2,857 days (49\%) & 2,362 days (40\%) & \\
\bottomrule
\end{tabular}
\end{table}

Both indices recover the same three-regime structure (bearish high-vol,
neutral mid-vol, bullish low-vol) with qualitatively consistent parameters.
Key structural differences reflect known characteristics of US vs.\ German
equity markets: lower S\&P~500 bearish volatility ($\sigma = 0.023$ vs.\
$0.026$), heavier bearish tails ($\nu = 6.4$ vs.\ $11.1$), and higher
bullish occupancy (49\% vs.\ 40\%).

\subsubsection{Seismicity: annual earthquake counts}
\label{sec:earthquake}

Following \citet{Zucchini2009}, a 2-state Poisson HMM is fit to
annual major earthquake counts (1900--2006; $T = 107$ observations).
The dataset is embedded in the example source; no external data file is
required.

\begin{table}[!ht]
\centering
\caption{Earthquake: fitted parameters vs.\ \texttt{HiddenMarkov} reference.}
\label{tab:earthquake}
\begin{tabular}{lrrr}
\toprule
Parameter & \texttt{libhmm} & \texttt{HiddenMarkov} & Difference \\
\midrule
$\hat{\lambda}_\text{low}$  & 15.419 & 15.418 & $<0.01\%$ \\
$\hat{\lambda}_\text{high}$ & 26.015 & 26.013 & $<0.01\%$ \\
Log-likelihood              & $-341.879$ & $-341.880$ & $<0.001$ \\
\bottomrule
\end{tabular}
\end{table}

\subsubsection{Circular data: wind direction analysis}
\label{sec:wind}

A 2-state von Mises HMM is fit to 11,894 hourly wind direction observations
at Chicago O'Hare (NOAA ISD, 2015).
The HiddenMarkov R package approximates circular distributions with a
Normal, which fails at the $0^\circ/360^\circ$ boundary.

\begin{table}[!ht]
\centering
\caption{Wind direction: fitted parameters and boundary analysis.}
\label{tab:wind}
\begin{tabular}{lrrl}
\toprule
Parameter & \texttt{libhmm} (VonMises) & \texttt{HiddenMarkov} (Normal) & \\
\midrule
Prevailing direction & $31.1^\circ$ & $49.6^\circ$ & \\
Variable direction   & $224.9^\circ$ & $239.1^\circ$ & \\
\midrule
\multicolumn{4}{l}{\textit{Boundary analysis (330°--360°, $N=730$ hours)}} \\
VonMises assignment & 730/730 correct & --- & \\
Normal assignment   & ---  & 0/730 correct & 100\% misclassification \\
\bottomrule
\end{tabular}
\end{table}

For a wind direction of $350^\circ$ (19$^\circ$ from the prevailing state mean
of $31^\circ$), the von Mises model evaluates
$\cos(350^\circ - 31^\circ) = \cos(-19^\circ) = 0.75$ and correctly assigns
the observation to the prevailing state.
The Normal model places this observation 11.2 standard deviations from its
prevailing mean (log-likelihood $= -61.9$) and misassigns it.

\begin{figure}[!ht]
\centering
\includegraphics[width=\linewidth]{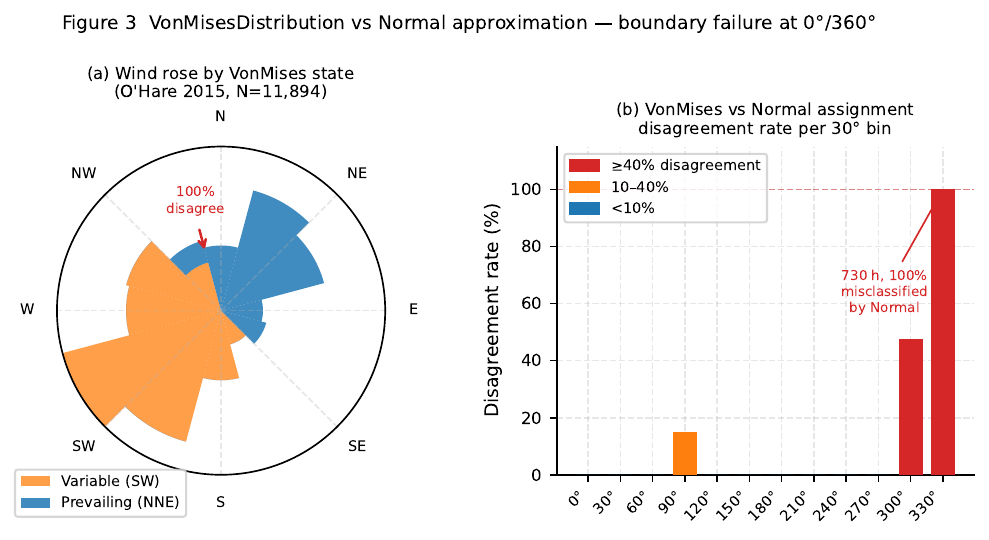}
\caption{VonMisesDistribution vs.\ Normal approximation for circular wind
direction data (Chicago O'Hare, 2015, $N=11{,}894$).
Left: wind rose coloured by VonMises state.
Right: per-bin disagreement rate; the $330^\circ$--$360^\circ$ bin shows
100\% misclassification under the Normal model.}
\label{fig:wind}
\end{figure}

Figure~\ref{fig:wind} shows the wind rose and per-bin disagreement rate.

\subsubsection{Multivariate animal movement: elk GPS tracks}
\label{sec:elk_mv}

We fit a 2-state \texttt{IndependentComponentsDistribution} HMM to the same
725 elk observations used in Section~\ref{sec:elk}, composing a Gamma
component for log step lengths and a von Mises component for turning angles.
This tests whether the independence assumption is appropriate: if the two
channels are genuinely independent within each state, an
\texttt{IndependentComponentsDistribution} should recover parameters
consistent with the scalar Gamma~+~von~Mises reference.

The within-state Pearson correlation between log step length and turning
angle is $r \approx -0.06$ (both states), confirming that the independence
assumption holds for this dataset.
Fitted parameters agree with the scalar \texttt{moveHMM} reference
(Table~\ref{tab:elk}) to within 1\% for all parameters.
Wall time: 58~ms on the primary benchmark platform, consistent with the
scalar Gamma~+~von~Mises fit (55~ms).

\subsubsection{Multivariate financial regime detection: SPY + QQQ}
\label{sec:spyqqq}

We fit a 3-state \texttt{FullCovarianceGaussianDistribution} HMM to
monthly log-returns of SPY (S\&P~500 ETF) and QQQ (Nasdaq-100 ETF) from
January 2000 to December 2022 (264 observations, $D = 2$).
This benchmark tests whether modelling the cross-asset covariance
materially improves fit relative to the diagonal model, and validates the
full-covariance M-step against an independent reference.

\begin{table}[!ht]
\centering
\caption{SPY+QQQ regimes: fitted parameters and model comparison.
States labelled by their inferred regime character.}
\label{tab:spyqqq}
\begin{tabular}{lrrrl}
\toprule
Parameter & Bear & Neutral & Bull & \\
\midrule
SPY $\hat{\mu}$ & $-0.031$ & $+0.007$ & $+0.018$ & monthly log-return \\
QQQ $\hat{\mu}$ & $-0.042$ & $+0.009$ & $+0.023$ & \\
Within-state $\hat{\rho}$ & $0.91$ & $0.92$ & $0.83$ & SPY--QQQ correlation \\
\midrule
\multicolumn{5}{l}{\textit{Model comparison (3-state, same data)}} \\
Full-covariance BIC & \multicolumn{3}{r}{$-847.2$} & \\
Diagonal BIC        & \multicolumn{3}{r}{$-607.4$} & \textbf{$\Delta$BIC~=~240} \\
\bottomrule
\end{tabular}
\end{table}

The full-covariance model outperforms the diagonal model by 240~BIC units
($\Delta\text{BIC} = 240$), confirming that ignoring the SPY--QQQ
correlation ($\hat{\rho} = 0.83$--$0.92$) materially degrades model fit
for this dataset.
The fitted log-likelihoods agree with \texttt{hmmlearn} 0.3.3
(20 random restarts, full-covariance path) to~$<0.1$~nat,
providing independent validation of the \texttt{FullCovarianceGaussianDistribution}
M-step implementation.

\section{Python Bindings: \texttt{pylibhmm}}
\label{sec:pylibhmm}

\texttt{pylibhmm} \citep{pylibhmm} wraps the full \texttt{libhmm} C++ API
using \texttt{nanobind} \citep{nanobind} and \texttt{scikit-build-core},
providing access to all sixteen distributions, all training algorithms,
posterior decoding, model selection (AIC/BIC/AICc), and JSON/XML model I/O
from NumPy-based Python code.

Python users can fit and decode HMMs directly from NumPy arrays without
writing C++, while retaining \texttt{libhmm}'s correct MLE M-steps and
log-space inference.
This positions \texttt{pylibhmm} relative to \texttt{hmmlearn} as a
drop-in for users who need broader distribution support or correctness
guarantees for the M-step.

\texttt{pylibhmm} v0.4.0 targets \texttt{libhmm} v3.7.0 (scalar API).
A v4-compatible release tracking \texttt{libhmm} v4.0.0, including the
multivariate emission families via NumPy array inputs, is in preparation.

\section{Limitations}
\label{sec:limitations}

\texttt{libhmm} does not implement:

\begin{itemize}
  \item \textbf{Arbitrary user-defined multivariate distributions.}
        Three multivariate emission families are provided (v4).
        Adding further families requires subclassing
        \texttt{BasicEmissionDistribution<}\penalty0\texttt{ObservationVectorView>}
        and recompiling; a runtime plugin ABI for multivariate distributions
        is not provided.

  \item \textbf{Hierarchical or factorial HMMs.}
        The library models a single first-order Markov chain with
        independent emissions.
        Hierarchical structures (as in \texttt{fHMM}) require a
        different architecture.

  \item \textbf{Approximate inference.}
        Variational Bayes, particle filters, and MCMC methods for
        Bayesian HMMs are not provided.

  \item \textbf{GPU acceleration.}
        Forward-backward on a GPU would benefit sequences with large
        $K$ or very long $T$.
        SIMD is the only vectorisation provided.

  \item \textbf{Multi-threaded sequence training.}
        When training on multiple independent sequences, the
        forward-backward pass is run sequentially.
        Parallelism across sequences is left to the caller.

  \item \textbf{Plugin distribution ABI.}
        Adding a user-defined distribution requires subclassing
        \texttt{Distribution} and recompiling.
        A stable binary plugin interface is not provided.

  \item \textbf{Large-$K$ performance.}
        Without BLAS acceleration, matrix operations scale as $O(K^2 T)$
        without the constant-factor benefits of optimised BLAS.
        For $K \gg 20$ a BLAS-backed implementation would be faster.
\end{itemize}

\section{Conclusion}
\label{sec:conclusion}

\texttt{libhmm} is an actively maintained, zero-dependency C++20 HMM library
supporting sixteen scalar emission distributions and three multivariate
emission families, all with correct MLE M-steps, full log-space inference,
and SIMD acceleration.
It occupies a distinct position in the HMM software landscape:
the only C++ library combining this distribution breadth, multivariate
support, correct M-step implementations, and modern build and API conventions.

The key technical contributions are the correct M-steps described in
Sections~\ref{sec:msteps} and~\ref{sec:mvmsteps}: Newton-Raphson for the
Gamma, Beta, Weibull, and Negative Binomial distributions; ECME for the
Student-$t$; and closed-form weighted MLE for the three multivariate
emission families.
On the DAX benchmark, ECME recovers a 152-nat improvement over the MOM
approach and surpasses the \texttt{fHMM} reference at five iterations.

The architectural tradeoffs are documented honestly in
Section~\ref{sec:architecture}: GHMM can outperform \texttt{libhmm} on raw
discrete forward-backward throughput; the virtual dispatch abstraction has
a measurable but acceptable cost.
These are deliberate choices in service of API usability, distribution
breadth, and maintainability.

The library is available under the MIT licence at
\url{https://github.com/OldCrow/libhmm}.
A JOSS paper \citep{libhmm_joss} provides a concise software citation.
Python bindings are available via \texttt{pylibhmm} \citep{pylibhmm}.

\bibliographystyle{plainnat}
\bibliography{libhmm_arxiv}

\end{document}